\documentclass[a4paper,amsmath,amssymb]{jpconf}
\usepackage{graphicx}
\usepackage{amsmath}

\newcommand{\bx}{\mathbf{x}}
\newcommand{\bq}{\mathbf{q}}
\newcommand{\boldf}{\mathbf{f}}
\newcommand{\hpsi}{\hat{\psi}}
\newcommand{\la}{\lambda}

\newcommand{\be}{\begin{equation}} 
\newcommand{\ee}{\end{equation}} 
\newcommand{\bea}{\begin{eqnarray}} 
\newcommand{\eea}{\end{eqnarray}} 
\newcommand{\bc}{\begin{center}} 
\newcommand{\ec}{\end{center}}

\begin{document}

\title{Spectral renormalization group theory on networks}

\author{Eser Ayg\"un$^1$ and Ay\c se Erzan$^2$ }

\address{
$^1$Department of Computer Engineering, and 
$^2$Department of Physics Engineering,\\ Istanbul Technical University, 
Maslak, Istanbul 34 469, Turkey
}
\ead{erzan@itu.edu.tr}

\begin{abstract}
Discrete amorphous materials are best described in terms of arbitrary networks which can be embedded in three dimensional space. Investigating the thermodynamic equilibrium as well as non-equilibrium behavior of such materials around second order phase transitions call for special techniques.

We set up a renormalization group scheme by expanding an arbitrary scalar
field living on the nodes of an arbitrary network, in terms of the
eigenvectors of the normalized graph Laplacian. The renormalization
transformation involves, as usual, the integration over the more ``rapidly
varying" components of the field, corresponding to eigenvectors with larger
eigenvalues, and then rescaling. The critical exponents depend on the
particular graph through the spectral density of the eigenvalues.

\end{abstract}


\section{Introduction}
\label{sec:Introduction}
Networks are ubiquitous in modeling discrete random media. For example granular media are characterized by contact or force 
networks~\cite{Mehta1,Mehta2}. Networks of slow regions are found to percolate near the glass transition.~\cite{Bennemann,Yilmaz} 
Apart from interest in the network properties themselves, networks provide a convenient scaffolding on which one may describe the fluctuations in quantities that do not live on periodic lattices or are not necessarily embedded in a metric space. 

This paper describes an attempt at developing a renormalization 
group treatment of a scalar field living on a complex network.

Critical phenomena on complex networks have been thoroughly studied. The field has been expertly reviewed by Dorogovsev et al.~\cite{dorogovtsev3}. The Bethe approach gives rise to exact results for networks which are asymptotically tree like~\cite{dorogovtsev1,dorogovtsev2}. Mean field like methods have been adopted to heterogeneous graphs~\cite{Bianconi}.   A Landau-type phenomenological approach has been developed for arbitrary networks~\cite{Goltsev_Landau}, where the explicit dependence of the critical behavior on the full degree distribution $P(k)$ is demonstrated. It is known~\cite{dorogovtsev1,dorogovtsev2} that for networks with a divergent $\langle k^2  \rangle$,  there is no phase transition except at infinite temperature in the ``thermodynamic limit." The phase transition exists and  is of infinite order for finite number of nodes $N$, and a special universal behavior is observed for all known models falling into this class. The critical behavior of the response function associated with the order parameter is mean field  (exact) for scale free networks having degree distributions with convergent second moments~\cite{dorogovtsev1,crit_fluct_Bianconi}.  
Exact real-space transformations  have also been performed on scale free hierarchical networks, see, e.g.
~\cite{Andrade_RSRG}. A Berezinskii-Kosterlitz-Thouless phase with power 
law correlations, and a phase transition that is infinite order has been 
observed on networks with strongly inhomogeneous degree 
distributions, e.g., graphs with degree distributions $P(k)\sim k^{-\gamma}$ and $\gamma$ 
sufficiently small~\cite{dorogovtsev2,Bauer1,Bauer2,Costin}, leading to departures from mean field behavior. 
 The presence of nodes with a diverging number of edges gives rise to 
Griffiths sigularities~\cite{Griffiths,GriffithsRSRG,Hinczewski_Griffiths}.

In spite of the rich literature on critical phenomena on arbitrary 
networks, the possibility of developing a ``field theoretic" 
renormalization group on networks has not been explored so far, at least 
within a condensed matter context. Although some basic mathematical 
concepts needed to construct such a method are available, it turns out that  there are also a 
number of unexpected challenges.

The paper is organized as follows. In Section \ref{sec:mathintro} 
some basic mathematical tools are assembled, in Section \ref{sec:FTRG}  
 the field theoretic renormalization group, \`{a} la 
Wilson is very briefly reviewed. In Section \ref{sec:NWFTRG} a 
renormalization group transformation is introduced  for the relevant Landau-Ginzburg 
Hamiltonian density on a scale free network and  the main 
difficulties are outlined. In section 5, a replica approach is used to perform quenched averages over stochastic spectral densities and a brief discussion of renormalization on deterministic hierarchical lattices is provided. In Section \ref{sec:conclusions} we present some conclusions and pointers for future work.
 
\section{The Graph Laplacian and the Laplace Spectrum}
\label{sec:mathintro}

Let us define a graph as a collection of $N$ vertices, or ``nodes," labeled $i=1,\ldots,N$ and connected to each other by edges.    The adjacency matrix  ${\bf A}$ completely specifies such a graph, with 
\be A_{ij}\equiv 
\begin{cases} 1 & {\rm if } \;\;(i,j)\;\; {\rm connected}\;\;{\rm by} \;\;{\rm an} \;\;{\rm edge} \\ 
0 & {\rm otherwise }  \end{cases} 	
\ee
If no self-interactions are allowed $A_{ii}=0$ . The ``degree" $k_i$ of the $i$th node is the number of edges that connect it to other nodes, which are then called its  ``neighbors," and $k_i=\sum_j A_{ij}$.
 
For an undirected graph, ${\bf A}$ is symmetric, by definition; in this paper we will deal only with undirected graphs.  The invariants of the adjacency matrix are of course independent of the order in which the nodes are labeled, and so are all the different statistics on the graph in which one may be interested, such as the degree distribution (the probability distribution of the number of edges that the nodes have). On the other hand, unless the nodes are identified in the same way, it is not trivial to compare two graphs.

We will focus here on graphs that are not necessarily embedded in metric spaces, so that the edges do not carry information regarding the ``distance" between neighboring nodes. Distance between an arbitrary pair of nodes $(i,j)$ on the graph is simply defined as the least number of edges one has to traverse in going from node $i$ to node $j$. 

The formal analogue of the Laplace operator on an arbitrary graph, is the 
graph Laplacian~\cite{Chung,Luxburg,Banerjee} 
 which is defined as
\be
{\bf L} \,=\, {\bf D}- {\bf A}
\ee
where 
\be D_{ij}\, = \delta_{ij} k_i
\ee
and $k_i$ is the degree of the $i$th node.

Note that for any scalar field  ${\bf f} = (f_1,\ldots, f_N)$ on a network of size $N$, 
\be
 ({\bf L} \boldf)_i\,=\, \sum_j L_{ij} f_j = k_i f_i - \sum_{j\in {\cal N}_i} f_j\;\;, \label{eq:graphlapl}
\ee
where ${\cal N}_i$ is the set of neighbors of the node $i$.
 
By construction, $\sum_j L_{ij} = 0$.  For an undirected graph  the eigenvectors
\be
{\bf L}\,{\bf u}_\lambda=\lambda {\bf u}_\lambda
\ee 
are orthogonal since ${\bf A}$ is symmetric, the eigenvalues $\la$ are real and non-negative (please do not  mistake the  symbol $\la$ for a wavelength!). If the graph consists of one connected component, $\lambda_0=0$ is non-degenerate. There is a finite gap between the zeroth  and the first eigenvalue $\lambda_1 >0$, which is bound away from zero for finite graphs.  It is convenient to put the eigenvalues in ascending order with $ 0< \lambda_1 \le \lambda_2\ldots \le \lambda_{N-1}$, and define the largest eigenvalue as $\lambda_{N-1} \equiv \Lambda$.

The following useful properties are easy to show. The elements of the eigenvector belonging to a non-degenerate zero eigenvalue will be constant (which can be chosen $c=N^{-1/2}$ for normalization),
\be
u_{\lambda_0}(i) =  c \;\; \forall i\;\;.
\ee
If the graph consists of a number of connected components, the matrix ${\bf L}$ is of block diagonal form; the number of zero eigenvalues is equal to the number of connected components, and the corresponding eigenvectors each have elements that are constant over one of the connected components, and zero otherwise.
The eigenvectors for  $\lambda \ne 0$, on the other hand, satisfy
\be
\sum_i {u}_\lambda(i) = 0 \;\;.\label{eq:zerosum}
\ee
However, it should be noted that, unlike plane waves, in general $\sum_\la   
u_\la(i)\, u_\la(j) \ne \delta_{ij}$, and similarly,  $ \sum_\la 
u_\la(i)\ne 0$. 

On a hypercubic lattice of dimension $d$,  the degree of each node $i$ is $k_i\,=\,2d$, 
and Eq. (\ref{eq:graphlapl}) is simply 
the second order difference operator acting on $\boldf$. For $d=1$, it is 
easy to see that if we divide the right hand side with the Euclidian 
distance between the neighboring nodes, $-\,{\bf L}$ goes over to the Laplacian 
in the continuum limit, i.e.,

\be -\nabla^2 f(x) = \lim_{\Delta\to 0} \{2f(x) - [f(x+\Delta) +
f(x-\Delta)]\}/\Delta^2\;\;.
\ee
Arguments for the convergence to continuous Laplace operators can be made rigorous~\cite{Chung,Luxburg}. Moreover~\cite{Luxburg},

\be {\bf f}^\dagger {\bf L} {\bf f} = {1\over 2} \sum_{i,j}^N A_{ij} (f_i-f_j)^2 
\label{eq:fLf} \ee 
converges to 
\be (f(x), {\cal L} f(x)) = \int_\Omega 
\vert \nabla f\vert^2 dx \,\, 
\ee 
where the integral is over the support of the function ${\bf f}$.

Note that the right hand side of Eq.(\ref{eq:fLf}) essentially counts 
the number of edges over which $(f_i-f_j)^2$ is appreciably different from 
zero.  One may use the Rayleigh-Ritz theorem~\cite{planemath},
\be \min_{||\boldf|| \ne 0} {{\bf f}^\dagger L {\bf f} \over {\bf 
f}^\dagger {\bf f}} = \min{\lambda}
\ee
where $\lambda$ are the eigenvalues of the operator ${\bf L}$, to partition the 
graph into two  clusters  which are connected by the fewest 
possible edges~\cite{Luxburg}. 
The non-trivial solution to the minimization 
problem is clearly found by setting ${\bf f}={\bf u}_{\lambda_1}$ 
and ${\min} \la=\lambda_1$. It can 
be verified by explicitly (by computing the eigenvector 
${\bf u}_{\lambda_1}$) that its elements  fall roughly into two comparable 
sized 
groups of positive and negative values. This clustering scheme can be 
generalized to larger numbers of clusters~\cite{Luxburg}.

So far, it looks like the graph Laplacian provides us with an operator whose eigenvectors will be the analogue of the complex exponentials, so that we will be able to perform a transformation on our function $\boldf$ and analyze fluctuations at different resolutions, in analogy with Fourier components of different wavenumbers. (As noted above, however, the kernel of the transformation  is not self adjoint, unlike the Fourier transform.) 

\begin{figure}[ht]
\begin{center}
\includegraphics[width=7.9cm]{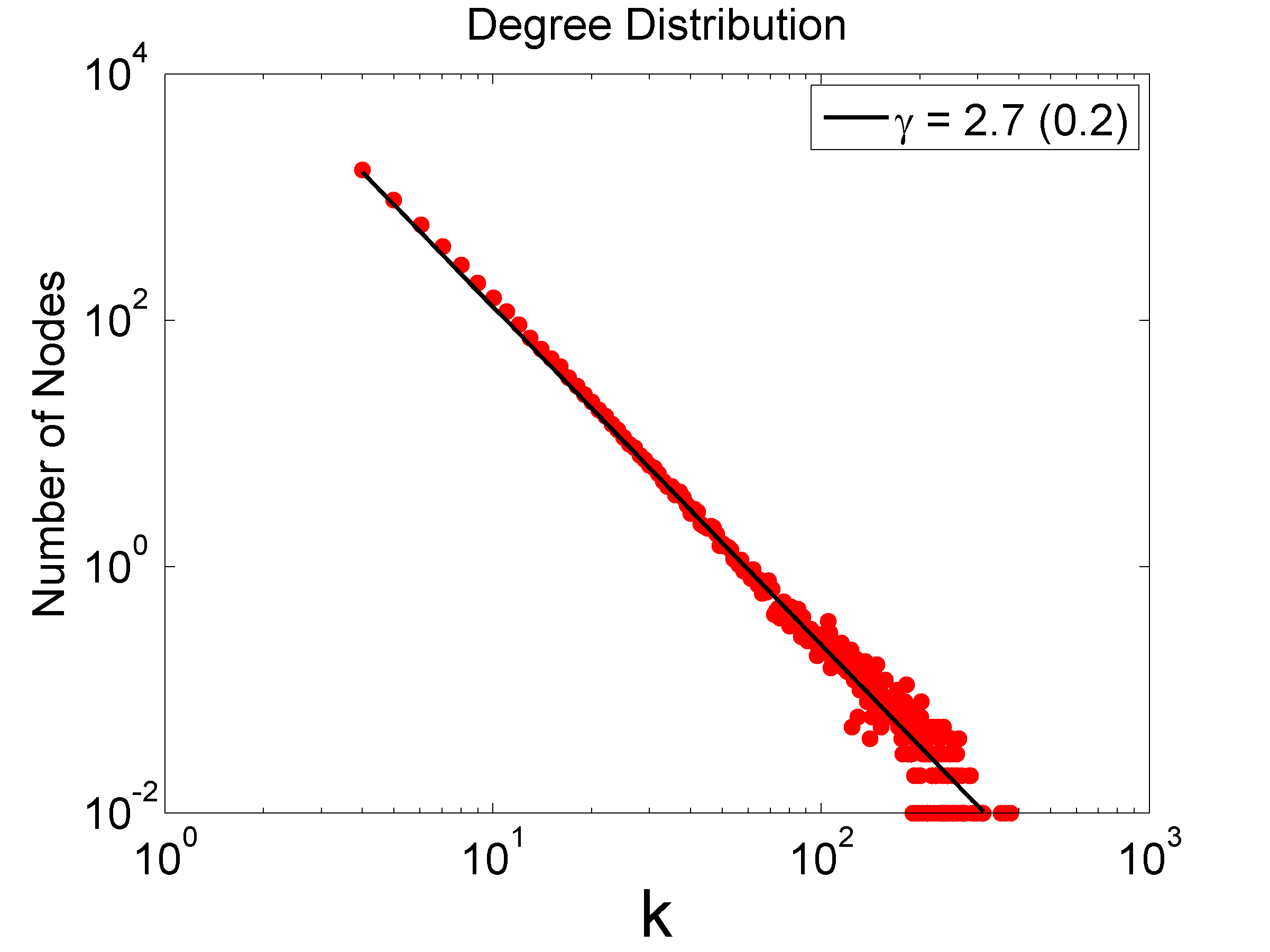}
\includegraphics[width=7.9cm]{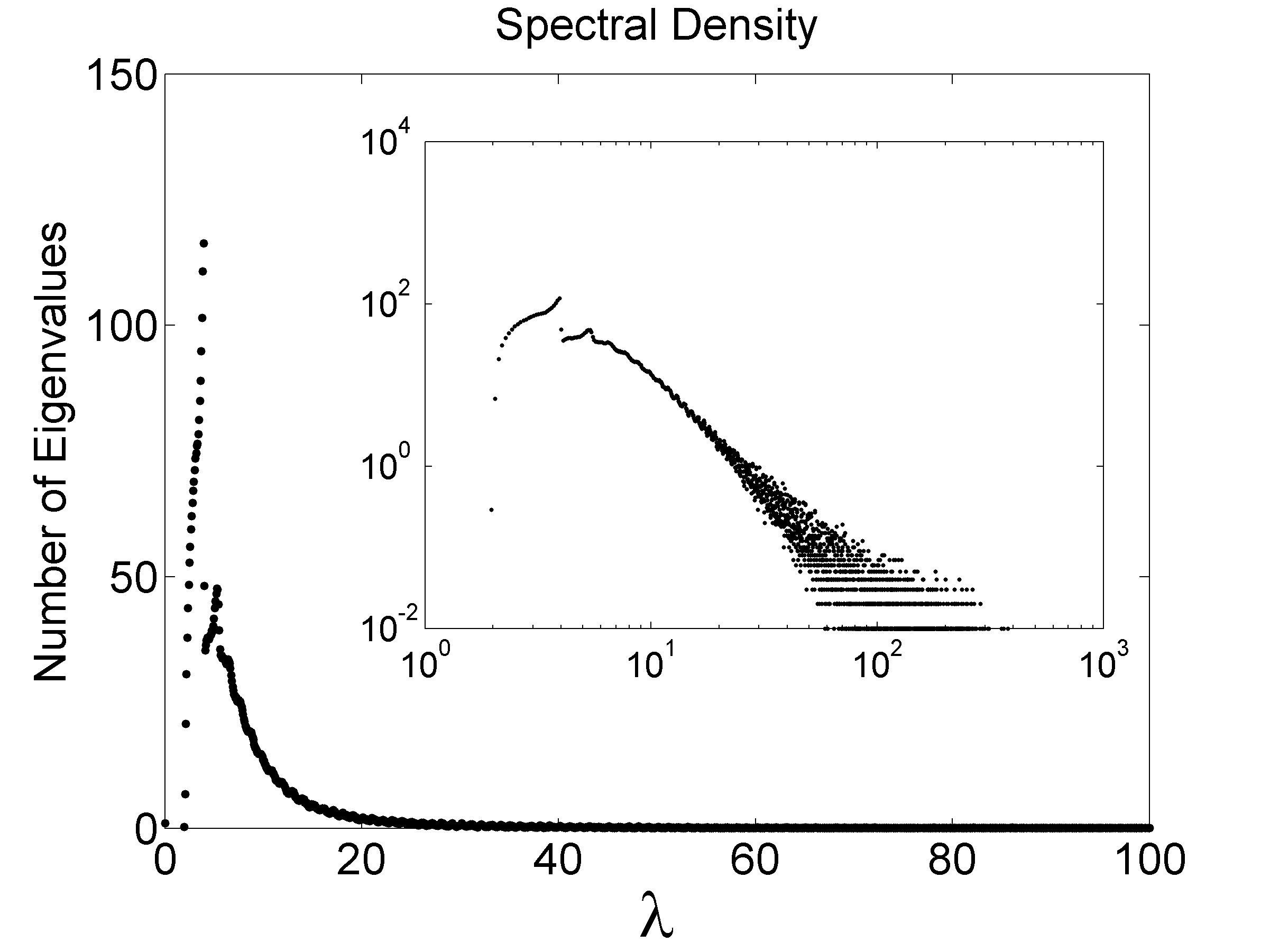}
\end{center}
\caption{\footnotesize  (Color on line.) The degree distribution (left panel) and Laplacian 
spectral density plotted against the eigenvalues $\lambda$  (right panel), 
for a ``scale free" network. Shown is the preferential attachment model\cite{Barabasi} for the number of initial nodes $m_0=5$, and $m=4$ new nodes added at each time step, grown to a size of $N=5000$ and averaged over 100 realizations.  The left panel has double logarithmic axes, and the degree distribution has been fitted to a power law $k^{-\gamma}$, with the root mean square error of the fit reported in parentheses (see Table~1). For this figure the (linear) $k$ axis has been binned into 5000 intervals. The Laplace spectrum (right panel) is shown both with linear and  double logarithmic (inset) axes. The van Hove singularity in the spectral density and the  crossover from an exponential to a power law decay is quite evident. Both histograms have been obtained with 5000 bins on the linear scale. See text.}
\label{fig:BAddLS}
\end{figure}

\begin{figure}[ht]
\begin{center}
\includegraphics[width=7.9cm]{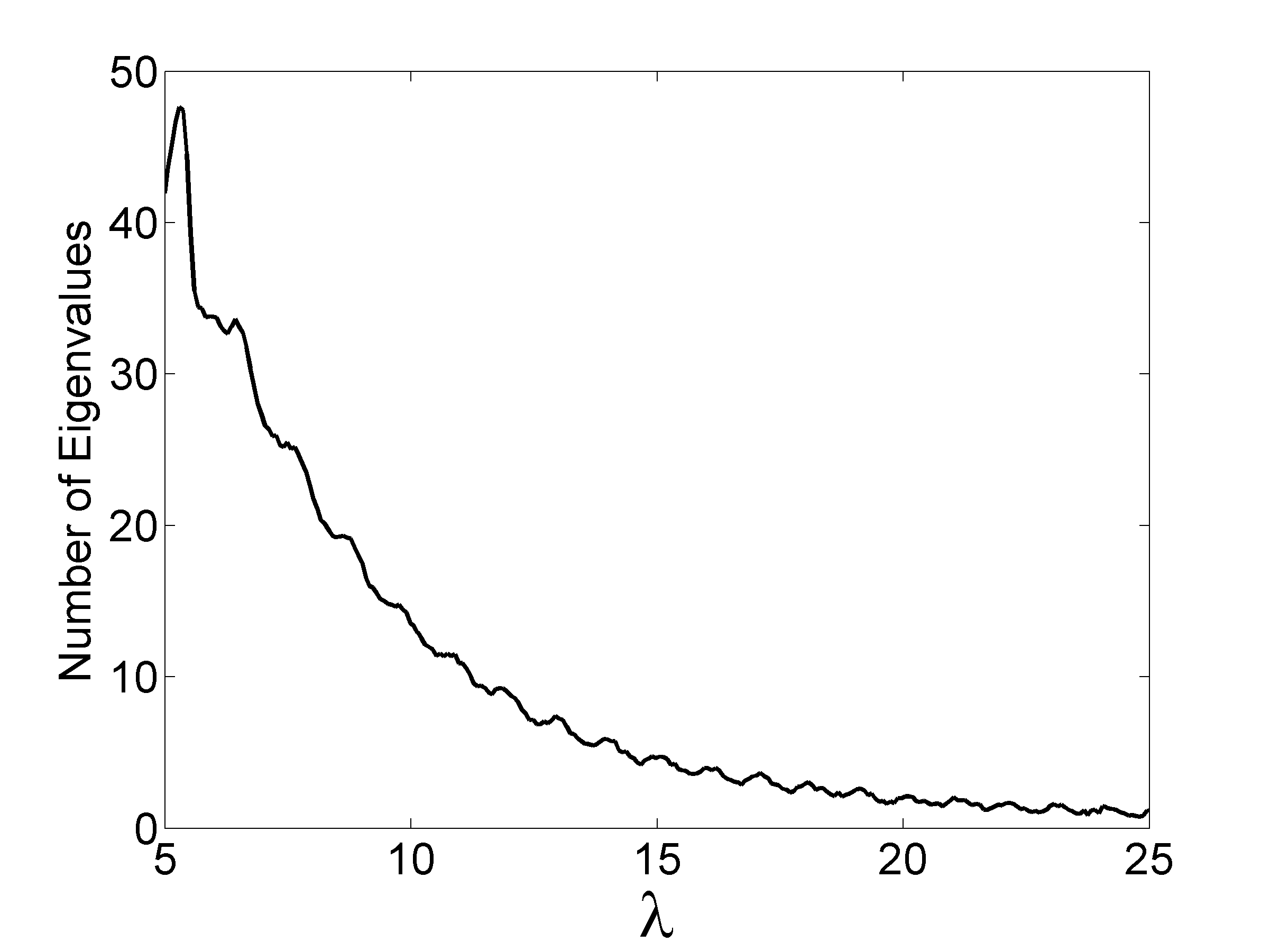} 
\includegraphics[width=7.9cm]{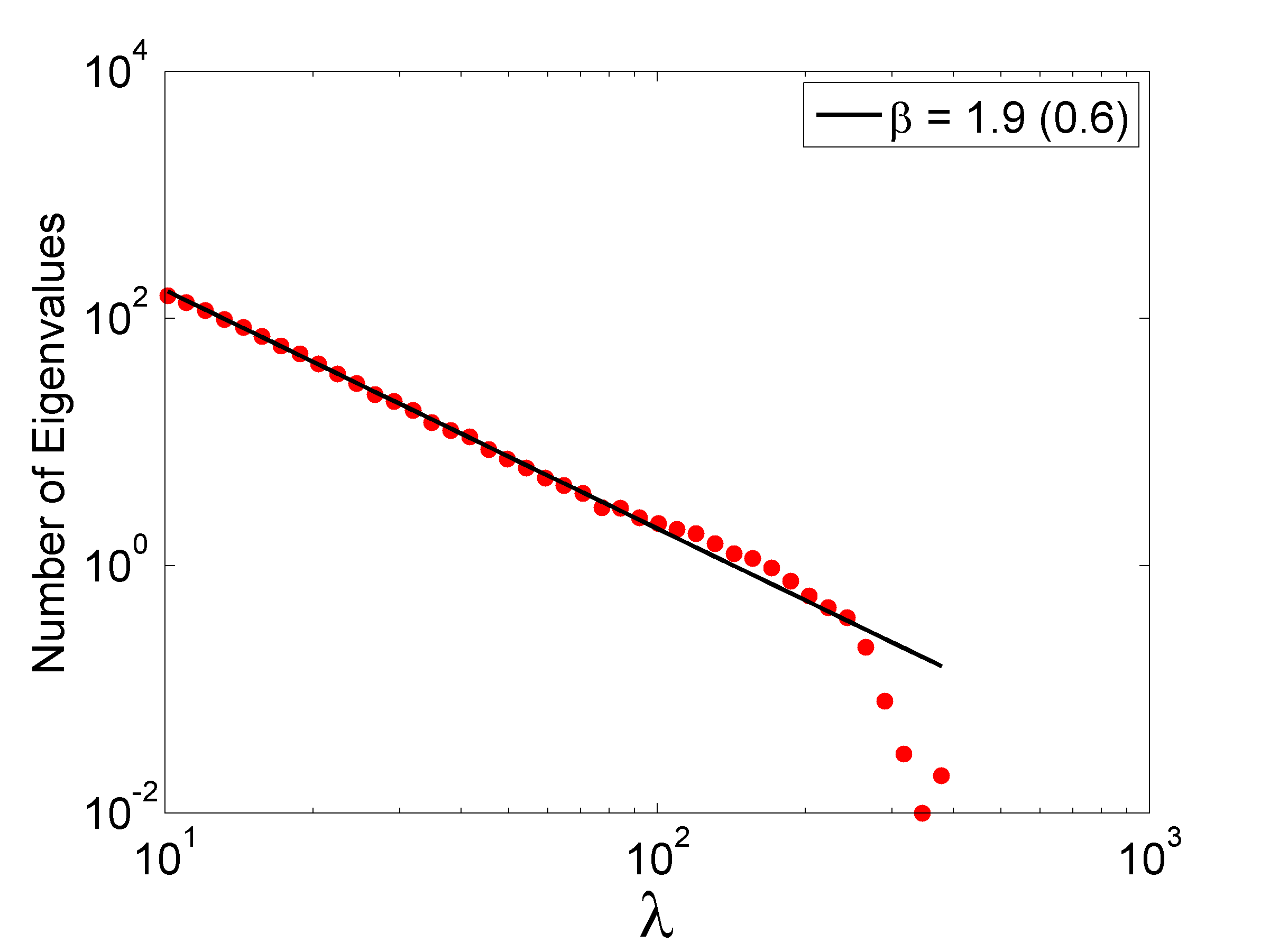} 
\end{center}
\caption{\footnotesize (Color online.) The tail end of the 
Laplace spectral density, ensemble averaged  (over 100 realizations) 
 for the same scale free network 
 as shown in Fig.~\ref{fig:BAddLS}. The left panel is on a  linear scale;  the points 
have been connected by a line.  The right panel shows a fit to 
$\rho(\lambda) \sim \la^{-\beta}$  on a log-log scale, with log-binning (see Table~1). 
}
\label{fig:BALStail}
\end{figure}

It is useful to take a look at the Laplace spectra of some sample networks.  Clearly, a complex, scale free network would be of greater interest in this context.  The degree distribution of a network generated using the Barabasi-Albert preferential attachment model~\cite{Barabasi} is shown in 
Fig.~\ref{fig:BAddLS}a. Although the degree distribution  fits a power law 
$\sim k^{-\gamma}$ perfectly over the whole of its range, the Laplace spectral density (Fig.~\ref{fig:BAddLS}b) is  not  scale free; it displays a sharp peak followed by a discontinuity and then, in the upper part of its range, it crosses over from an exponential to a power law decay, $\rho(\la)\sim \la^{-\beta}$, followed by a sharp cutoff (see Fig.~\ref{fig:BALStail}b). The behavior is very reminiscent of the density of vibrational states of a solid, exhibiting the van Hove singularity~\cite{IP1,IP2}.

\begin{table}[ht]
\begin{center}
\caption{Numerical values for $\gamma$ and $\beta$  for the Barabasi-Albert\cite{Barabasi} model. The number of initial nodes 
is $m_0=5$; the number of nodes added at each time step is $m$, and the 
total number of nodes is $N=5000$. The ensemble averages of the degree and 
spectral distributions over 100 realizations have been fitted, and these 
numbers are reported below. For the reported exponents, the $k$ and  $\lambda$ values have been log-binned into 500 and 50 bins respectively. The parentheses are root mean square error of the fits as illustrated in Fig. \ref{fig:BALStail}. }
\vskip 0.3cm
\begin{tabular}{l c c }
\hline
$m$ & $\gamma$ & $\beta$  \\
 \hline
3  & $2.7$ (0.05) & $1.9$ (0.03) \\
4  & $2.7$ (0.2$\;\;$)  & $1.9$ (0.06)\\
5  & $2.9$ (0.01) & $1.9 $ (0.09)\\
\hline
\end{tabular}
\end{center}
\vspace{-0.6cm}
\label{tb:table}
\end{table}

We do not have an 
analytic relation between $\beta$ and $\gamma$; it is quite possible that $\gamma$ all by itself 
does not determine $\beta$ and more work is needed to understand this 
dependence.

The ensemble averaged  tail of the Laplace spectrum is shown in Fig.~\ref{fig:BALStail}. 
The detail of Fig.~\ref{fig:BALStail} is astonishing.  It displays a 
self-similar piling up of further and further decorations superposed on the power law decay.  (For a BA 
model with aging of the sites~\cite{doro_aging}, the Laplace spectrum displays even more pronounced periodic modulations.) 
In fact, if these decorations can be represented as a 
self-similar Weierstrass-Mandelbrot function~\cite{Mandelbrot},
\be
W(\lambda) \sim W_0(\lambda) + \sum_{n=0}^N \frac{g(s^n \lambda)}{s^{\phi n}}\;\;\label{eq:WMfunction}
\ee 
where $g$ is a periodic function and  $s>1$ is a  constant, then $W(\la)$ has the scaling form  
\be
W(s\lambda) \sim s^\phi W(\la)\;\;. \label{eq:scalingform1}
\ee

\section{Order-Disorder transitions, fluctuations and the Field Theoretical Renormalization Group}
 \label{sec:FTRG}
The Ising model, with a scalar field $\sigma_i = \pm 1$ residing on the vertices of a lattice, is the basic paradigm for order-disorder transitions in statistical physics. The Ising Hamiltonian, or energy function for any given set $\{\sigma_i\}$, is,  for short range interactions, 
\be
H= -J \sum_{<ij>} \sigma_i \sigma_j -h \sum_i \sigma_i
\ee
where $<ij>$ signifies that $i$ and $j$ are connected directly by an adge 
on the lattice.  
For $J>0$, configurations in which the ``spins" 
$\sigma_i$ 
predominantly have the same sign as their neighbors lower the energy with 
respect to random configurations.  In thermal equilibrium at sufficiently 
low temperatures and sufficiently high connectivity of the network (for 
the embedding dimension $d\ge 2$ on periodic lattices), the spins become 
``ordered," i.e., the ``magnetization"  which is equal to the 
thermodynamic expectation value $\langle \sum_i \sigma_i \rangle$ becomes non-zero,, 
even when the external magnetic field $h$ is equal to zero.  The magnetization
can be regarded as an ``order parameter."  At the precise temperature 
$T_c$ where this ordering sets in, a continuous 
phase transition takes place. 

At 
the ``critical" point $T=T_c$, $h=0$, the system is characterized by 
fluctuations at all wavelengths and a diverging correlation length.  As a 
result, the system becomes invariant under scale transformations on a 
Euclidean lattice.  Slightly away from the critical point, changing the 
resolution at which the system is observed has the effect of changing the 
effective temperature (and magnetic field). The rate at which this change 
occurs determines the critical exponents which characterize the singular 
behavior of the free energy, order parameter and response functions of 
the system.~\cite{Goldenfeld,Kadanoff,Kardar} (In the rest of this section we mostly follow 
the presentation of Goldenfeld~\cite{Goldenfeld}, Chapter 12.)

\subsection{The Ginzburg-Landau Approach}

The Ginzburg-Landau approach~\cite{Kardar} to critical fluctuations 
involves a passage to a continuum description, which is based on a local 
averaging of the discrete variables such as $\{\sigma_i\}$ over volumes 
still very small compared to the total size of the system, so that one may 
eventually take them to be of infinitesimal size. In this way, a 
continuous variable corresponding to the local order parameter density is 
obtained.  Let us call this variable $\psi({\bx})$ where $\bx$ indicates 
the spatial coordinates.  The effective Ginzburg-Landau (G-L) Hamiltonian in the absence of an external magnetic field
is then given by,
\be
H = \int_\Omega d {\bx} \{\frac{1}{2} \left[ r_0 \psi^2(\bx) - 
\psi(\bx)\nabla^2\psi \right]+  v_0 \psi^4(\bx)\}\;\;.\label{eq:GL}
\ee
where the coupling $r_0$ is assumed to be proportional to $(T-T_c)/T_c \,\equiv\,t$, and $v_0\,>\,0$. It is customary to absorb  a factor of $1/k_BT$  into $H$, with $k_B$ being the Boltzmann constant, so that $H$ is dimensionless. This effective Hamiltonian can be expressed in terms of the Fourier coefficients of the order parameter, 
\be  \hpsi(\bq) = {1\over 2\pi} \int_\Omega e^{i\bq \cdot\bx} \psi (\bx) d\bx \;\;.
\ee
The $\bq $ signify wavevectors, the components of which are $q_\alpha = 2\pi n_\alpha /|\Omega|^{1/d}$, with each $n_\alpha$ ranging over the positive and negative integers (periodic boundary conditions are assumed). For a $d$-dimensional hypercubic domain $\Omega$, in the continuum limit, the density of wave vectors within a volume $d\bq $ is given by $|\Omega|/(2\pi)^{d}$. Note that $\bq $ are the eigenvalues of the gradient operator on the complex exponentials (which are the eigenfunctions); the eigenvalues of the Laplace operator on the same functions are simply $-|| \bq ||^2$. 

Let us write the effective Hamiltonian as $H\,=\,H_0 \,+\,H_{\rm int}$, 
where $H_0$ is quadratic and diagonal in the $\psi(\bq )$,
\be
H_0 =  \frac{1}{2}\int_{0}^\Lambda \frac{q^{d-1}dq}{(2\pi)^{d}} [r_0 
+q^2 ] 
\hpsi^\ast(\bq ) \hpsi(\bq )  \;\;. \label{eq:GL_fourier}\ee
Assuming isotropy in $q$-space, we have gone over to polar 
coordinates, obtaining the density (per unit volume $|\Omega|$) of 
wavevectors in the interval $dq$ to be 
$S_d \,\frac{q^{d-1}}{(2\pi)^{d}}$, where $S_d$ is the area of the unit 
sphere in $d$ dimensions. It is important to note that this
density is in the form of a power law in $q$. The upper cutoff $\Lambda$ is given by 
${2\pi/l}$ 
where $l$ is the lattice spacing. 

The interacting part of the Hamiltonian, $H_{\rm int}$, involves couplings between Fourier components at different wavenumbers (and therefore different spatial scales) and is given by,
\be H_{\rm int}= v_0 \int_{0}^\Lambda d\bq_1\ldots d\bq_4 \hpsi(\bq_1)\ldots \hpsi(\bq_4)\;\delta(\sum_{i=1}^4 \bq_i)\;\;.\label{eq:psi4}
\ee

The partition function is,
\be
Z= \int_{-\infty}^\infty \ldots \int_{-\infty}^\infty \prod_{0\le 
|\bq | 
\le \Lambda}  d\hpsi_\bq  e^{-H}\;\;. \label{eq:partfunc1}
\ee
\subsection{Renormalization Group \`{a} la Wilson}

Renormalization \`{a} la Wilson~\cite{Wilson} is carried out by {\it i)} 
integrating out the relatively small wavelength (large $q$) 
components of the fluctuating field  in the partition function~\cite{Goldenfeld,Kadanoff,Kardar}. Choosing a re-scaling parameter $b$, and dividing the range of $q$ into two parts, the partition function  can be written (exactly) as, 
\be
Z= \int_{-\infty}^\infty \ldots \int_{-\infty}^\infty \prod_{0\le 
|\bq | 
\le \Lambda/b}  d\hpsi_\bq ^{\rm lower} 
\prod_{\Lambda/b< |\bq | \le \Lambda} d\hpsi_\bq ^{\rm upper} e^{-H}\;\;. \label{eq:partfunc}
\ee
where we have added an extra label on the functions $\hpsi(\bq )$ to indicate the range (upper or lower)of wave vectors  within which $\bq $ lies.
We would now like to do the integration over the fluctuations in the upper $q$ range, $\hpsi_\bq ^{\rm upper} $. This can of course only be done exactly for $u_0=0$, and therefore the interacting part of the G-L Hamiltonian has to be taken care of perturbatively. 
{\it ii)} The next step involves rescaling  the $q$ to restore the 
original range of scales over which the fields fluctuate, and thus the original form of the Hamiltonian density.  {\it iii)} Requiring that the 
coefficient of the diffusive  ($q^2 $) term in $H_0$ remain invariant under this operation fixes the rescaling factors acquired by the fields $\hpsi(\bq )$ and the couplings $r$ and $v$.  It can easily be demonstrated that successive transformation with the parameter $b$ satisfy all the {\it semi}group properties (there is no inverse). 

The usefulness of the field theoretic approach of Wilson lies in the precise prescription for the Renormalization Group (RG) transformation, and the fact that although approximations have to be made for an interacting theory they can be systematically improved, as a power series in the coupling $v_0$, in contrast to the ``Real Space" RG approach~\cite{Goldenfeld}, where one has to perform uncontrolled approximate  partial summations over the partition function $Z$. Moreover, such concepts as an upper critical dimension, beyond which critical behavior is exactly mean field,  emerge only within the field theoretic RG formalism. 

\section{Field Theoretic RG on a Complex Network?}
\label{sec:NWFTRG} 

Having in mind the advantages of the Wilson approach to the 
renormalization group, we would like to construct the analogue of the 
field theoretic RG on complex networks. The idea is to expand order 
parameter fluctuations in eigenvectors of the graph Laplacian, write down 
the equivalent of a Ginzburg-Landau Hamiltonian, and then perform partial 
summations over the partition function, to eliminate the high-eigenvalue 
components.  The development will closely follow the prescription in the 
previous section.

\subsection{Order parameter expansion in eigenvectors of the Laplacian for a Gaussian model}

Let us model our approach on the Ginzburg-Landau expansion of the Hamiltonian 
as in Eq.~(\ref{eq:GL},\ref{eq:GL_fourier}) and initially take just a
Gaussian, or non-interacting model, with a field $\psi(i)$ residing on the vertices of a graph, with $i=1,\ldots,N$. In the absence of a field (which can be included without any problem)
\be
H_0 = \frac{1}{2} \sum_i^N  \psi(i)\,[ r_0 \,+\,L] \, \psi(i)  \;\;.
\ee
Let us define the transformed fields $\hpsi(\la)$ by 
\be \psi(i) =  \sum_{\la} \hpsi(\la)\, u_\lambda(i)\;\;,
\ee
in terms of ${\bf u}_\lambda$, the  normalized eigenvectors of the Laplace operator associated with the eigenvalues $\la$, and
\be \rho(\lambda) = {1\over N} \sum_i^N \delta(\lambda - \lambda_i)\;\;\label{eq:rho}
\ee
as the spectral density, i.e., the density of eigenvalues on the positive real line. Then we get,
\be
H_0=\frac{1}{2}\int_{0}^\Lambda \rho(\lambda)\, d\lambda \, \hpsi(\lambda)\,
[r_0 +\lambda]\, \hpsi(\lambda) \;\;. \label{eq:L_expansion}
\ee
Note that, being the eigenvalue of the Laplacian, $\la$ takes the place of the wavenumber squared, i.e., $q^2$  (it is actually dimensionless). 

\subsection{Naive renormalization of the Gaussian theory}
\label{ssec:naivegaussian}

The partition function can  now be written as an integral over the different coefficients $\hpsi_\lambda \equiv \hpsi(\la)$ (we will use these notations interchangeably). 
In order to perform the partial integration of the partition function as in  
Eq.~(\ref{eq:partfunc}), we choose a scale factor $s$.
We must now integrate out the coefficients $\hpsi_\lambda$ for $\lambda > \Lambda/s$, where $\Lambda$  is the largest eigenvalue of ${\bf L}$ on the given network.

Denoting the Gaussian partition function with $Z_0$, we have,
\be
Z_0= \int_{-\infty}^\infty \ldots \int_{-\infty}^\infty \prod_{0\le \lambda 
\le \Lambda/s}  d\hpsi_\lambda^{\rm lower} 
\prod_{\Lambda/s< \lambda \le \Lambda} d\hpsi_\lambda^{\rm upper} e^{-H_0}\;\;, \label{eq:Lap_partfunc}
\ee
where $H_0$ is now given by Eq.~(\ref{eq:L_expansion}).
Performing the set of integrals over the  $\hpsi_\lambda^{\rm upper}$ yields, 
\be
Z_0=  e^{g(r_0)} \int_{-\infty}^\infty \ldots \int_{-\infty}^\infty 
\prod_{0\le \lambda 
\le \Lambda/s}  d\hpsi_\lambda^{\rm lower} 
 e^{-H_0^{\rm lower}}\;\;, \label{eq:Lap_partf1}
\ee
where for future use we can define $ Z_0^{\rm upper} = e^{g(r_0)}$, 
where $g(r_0)$ is the free energy contributed by the 
high-$\lambda$ degrees of freedom.
Since  $ Z_0$ only involves doing Gaussian integrals, we easily calculate 
\be Z_0^{\rm upper} =
 \prod_{\Lambda/s < \la \le \Lambda} \sqrt{{2\pi\over \la+r_0}} \;\;.\ee

The Hamiltonian involving the remaining degrees of freedom is,
\be
H_0^{\rm lower} =\frac{1}{2} \int_{0}^{\Lambda/s} \rho(\lambda) \, d\lambda \, \hpsi_\la^{\rm lower} 
\, [r_0 +\lambda] \, \hpsi_\la^{\rm lower} \;\;. \label{eq:renormH1}
\ee 
The RG transformation will be complete when we rescale the $\lambda$ with $s$ so that the integral  is once again over the range $(0,\Lambda)$, and we can drop the label ``lower" from the Hamiltonian, while having to ``renormalize" the coupling $r_0$. 

Let us  make the naive assumption that  the spectral density is a homogenous function over its whole range, $\rho(\lambda) \sim \lambda^{-\beta}$, where  $\beta$ depends on the degree distribution, in particular on the exponent $\gamma$.  

Now  make a change of variables $\lambda^\prime = s \lambda$ (so that $\lambda = 
\lambda^\prime/s$).  Define the rescaling factor for the fields via $\hpsi^{\rm lower}(\lambda^\prime/s) = z \hpsi^\prime (\lambda^\prime)$. The resulting Hamiltonian is now going to be in the same form as Eq.~(\ref{eq:L_expansion}), except that $r_0$ will acquire a multiplicative factor.  Calling the renormalized  Hamiltonian $H_0^\prime$, 
Eq.~(\ref{eq:renormH1}) becomes,
\be
H_0^{\prime} =\frac{1}{2}\int_{0}^{\Lambda} \rho(\lambda^\prime) d\lambda^\prime s^{\beta-1} z^2\hpsi^\prime(\lambda^\prime) 
\left[r_0 +{\lambda^\prime \over s}\right] \hpsi^\prime(\lambda^\prime) \;\;\label{eq:renormH2}
\ee 

We assume we can fix the renormalization factor $z$ of the fields 
$\hpsi^\prime$ for all $\la^\prime$, by requiring the coefficient of the 
$\lambda^\prime$ term in the Hamiltonian to remain unchanged.  This gives 
$z=s^{(2-\beta)/2}$.  Substituting this back into Eq.~(\ref{eq:renormH2}) the 
various rescaling factors simplify and we find that we can define the 
renormalized coupling 
\be
r = s r_0 \;\;\label{eq:Gaussian_r}
\ee
Notice that as long as the spectral density is homogeneous, 
this result is inevitable for the Gaussian theory, independent of the scaling exponent $\beta$  for the spectral density. ($\beta$ should not be confused with the critical exponent of the order parameter!) Recall that the spectral density for the wavenumbers $\bq$ in the 
Euclidean case reviewed in Section~\ref{sec:FTRG}, was  $\propto q^{d-1}$. It is useful, for later comparison, to define $-\beta=D-1$. The fact that the spectral density decays, instead of growing with $\lambda$  is going to have strong consequences later on. 

In the Ginzburg-Landau approach we have assumed that $r_0 \propto t$, therefore the rescaling factor in 
front of $r_0$ is related to the temperature renormalization, i.e., 
we here have $t\to t^\prime = s^y t$, with $y=1$.

Recall that the eigenvalues of the Laplacian are ``like" $q^2$, so that the  
scaling factor of the ``length like" quantities in the problem is only $\sqrt{s}$. If we are very cavalier, we can identify 
the analogue of the correlation length exponent $\nu = (2y)^{-1} =1/2$, as we 
would expect from a Gaussian theory.

Since we have not embedded our network in a metric space, 
ascribing to $\lambda$ the dimensionality of a $({\rm length})^{-2}$, and 
similarly making the identification of $(2y)^{-1}$ with the correlation {\it 
length} exponent is rather tenuous, just as it is somewhat problematic to 
define a correlation length on this non-metric space. (Note that the dimension $d$ is also not defined for this system. However, for the product $d\nu$, the identification $2-\alpha = d\nu$ can still be made, where $\alpha$ is the specific heat exponent.)  

\subsection{The $\psi^4$ theory}

Consider adding an interaction term  $H_{\rm int} = v_0 \sum_i\psi^4 (i) $, to the non-interacting or Gaussian theory.  In terms of the transformed fields this gives, 
\be H_{\rm int}= v_0 \int_{0}^\Lambda \prod_{\mu=1}^{4} \left[ \rho(\la_\mu) d\lambda_\mu \right] \hpsi(\la_1)\ldots \hpsi(\la_4) \Phi(\la_1,\la_2,\la_3 \la_4 )\;\;.
\ee
where $\Phi(\la_1,\ldots, \la_4) \equiv \sum_{i=1}^{N} \prod_{\mu=1}^4 u_{\la_\mu}(i) $.

We now encounter a property of the eigenvectors of the ordinary Laplacian on a Euclidean lattice (e.g., plane waves)  which is not shared by the graph Laplacian: An element by element product of the eigenvectors  (eigenfunctions) of the ordinary Laplacian  yields yet another eigenvector since the exponentials add, modulo $2\pi$.
For the graph Laplacian, 
\be u_{\lambda_k}(i)u_{\lambda_l}(i)\ldots u_{\lambda_m}(i) \ne u_{\lambda_n}(i)\;\;.\label{eq:notdelta}
\ee
in general, i.e., no $\la_n$ can be found such that the equality holds.
If the equality were to hold, we would have had $\Phi=0$ unless $\la_n=0$, by 
Eq.~(\ref{eq:zerosum}). Recall that 
in the ordinary Euclidean case,  the $\psi^4$ term in the Fourier transform representation gives rise to a Dirac delta 
function connecting the wavevectors, $\delta(\bq_1+\bq_2+\bq_3+\bq+4)$, i.e., a sum rule on the total incoming and outgoing 
``momenta," Eq.~(\ref{eq:psi4}), while here, due to Eq.~(\ref{eq:notdelta}), no such 
rule applies.  This gives rise to a markedly 
different scaling behavior as we will see later on.

The interaction term has to be handled perturbatively, and following the 
same prescription~\cite{Wilson,Goldenfeld} here we get,
\bea
\lefteqn{Z(r_0, v_0)=  \int_{-\infty}^\infty \ldots \int_{-\infty}^\infty 
\prod_{0\le \lambda 
\le \Lambda}  d\hpsi_\la   e^{-H}} \nonumber \\
& & \mbox{} = Z_0^{\rm upper}  \int_{-\infty}^\infty \ldots 
\int_{-\infty}^\infty \prod_{0\le \lambda 
\le \Lambda/s}  d\hpsi_\la   e^{-H_0^{\rm lower}} \langle  e^{V[\hpsi^<, \hpsi^> ]} \rangle_0^{\rm upper}  \;\;.
\eea
where the interaction term now contains coefficients with $\la$ in both the lower and the upper range,

\be V[\hpsi^<, \psi^> ] \equiv -H_{\rm int} \left[ \hpsi(\lambda)^{\rm lower}, \hpsi(\lambda)^{\rm upper} 
\right]
\ee
 where we have implicitly defined  $\hpsi(\lambda)^>\equiv \hpsi(\lambda)^{\rm upper} $, etc.,  and the brackets mean,
\be 
\langle  e^{V[\hpsi^<, \psi^> ]}  \rangle_0^{\rm upper} = {1\over  Z_0^{\rm upper}}  
\int_{-\infty}^\infty \ldots \int_{-\infty}^\infty \prod_{\Lambda/s\le \la \le \Lambda}  
d\hpsi_\la   e^{-H_0^{\rm upper}} e^{V[\hpsi^<, \hpsi^> ]} 
\;\;.\label{eq:perturbation}
\ee

At this point, it is standard to take a cumulant expansion, 
$\langle e^{-x}\rangle \simeq e^{-\langle x\rangle} e^{{1\over 2}[\langle x^2 \rangle - \langle x\rangle^2]}$.
So now we have to compute $\langle V[\hpsi^<, \psi^> ]\rangle^{\rm upper}_0$ and $\langle V^2[\hpsi^<, \psi^> ] \rangle^{\rm upper}_0 $.
The angular brackets involve the same Gaussian measure as in Eq.~(\ref{eq:perturbation}) 
and the only terms that survive are those that 
are even in the $\hpsi(\lambda)^>$, with
\be
\langle \hpsi(\lambda)^{>} \hpsi(\lambda^\prime)^{>} \rangle_0^{\rm upper} 
= \delta_{\la\,\la^\prime} {1\over \la +r_0} \equiv \delta_{\la\,\la^\prime} G(r_0,\la)\;\;. \label{eq:green}
\ee
defining a ``contraction" between two fields  $\hpsi(\lambda)^{>}$ and $\hpsi(\lambda^\prime)^{>}$. 
Unless we refer specifically to the $r_0$ dependence, we will drop this argument and just write $G(\la)$ for the Green's function.
\subsection{First order in perturbation theory}

To keep track of these computations one makes use of Feynman diagrams, and the 
diagrams we need are given in Fig.~\ref{fig:feynmangraphs} in Appendix B. As usual, the 
contribution 
from 
$\langle V[\hpsi^<, \psi^> ]\rangle $ to the quadratic interaction comes from the singly contracted term, 
Fig.~\ref{fig:feynmangraphs}a, with the combinatoric  multiplicative factor 6.~\cite{Goldenfeld}. This term is given by
\be 
 Q_2^{(1)}= 6 v_0 \int_0^{\Lambda/s} d\la_1 d\la_2 \rho(\la_1) \rho(\la_2) \hpsi^< (\la_1)\hpsi^<(\la_2)  \sum_i I_1 (i)\,u_{\la_1}(i)u_{\la_2}(i) \;\;, \label{eq:1stOr}
\ee
where
\be
I_1(i) = \int_{\Lambda/s}^\Lambda\, d\la\, \rho(\la) G(\la) u_{\la}^2 (i)\;\;.\label{eq:I_1}
\ee
$I_1(i)$ does not depend any more on any of the eigenvalues, and therefore 
is a constant under re-scaling, however it depends on $r_0$, $s$, and 
$\Lambda$. (see Appendix B)

If, for the moment, we neglect the dependence of the integrand on the 
eigenvectors ${\bf u}_{\la_\mu}$, $\mu=1,2$, the scaling behavior of 
$Q^{(1)}_2$ is obtained by counting powers of $s$ which arise when we make 
the transformation to $\la^\prime = s\la$. This is the same as counting 
the number of integrals over $\la$ (each contributes a power $1-\beta$) 
and counting the powers of $\hpsi$, which contribute a factor of $z$ , 
i.e., a power of $(2-\beta)/2$ each.  In total we have the rescaling 
factor $-2+2\beta+ 2-\beta= \beta$. Taking into account the forefactor of 
$1/2$ appearing in $H_0$, Eq.~(\ref{eq:renormH1}) we get,
\be
 {1\over 2}Q_2^{(1)}= 12 s^\beta v_0  \int_0^{\Lambda} d\la_1 d\la_2 \rho(\la_1) \rho(\la_2) \hpsi (\la_1)\hpsi(\la_2) 
 \sum_i I_1 (i) 
u_{\la_1}(i)u_{\la_2}(i) \;\;, 
\ee
 for the contribution from the interaction term to the quadratic coupling, to first order in $v_0$. This contribution is not diagonal in the eigenvalues $\lambda$, unlike the original Gaussian coupling  in $H_0$, and moreover it 
depends on $\la$, so in principle the original $H_0$ also has to be modified. Note that, had the new 
quadratic term been diagonal, the renormalization factor would have been $s^{-1+\beta}z^2 = s$, rather than $s^\beta$.

Grouping together all the quadratic terms, absorbing any possible 
corrections to the scaling coming from $ \sum_i I_1 
(i)\,u_{\la_1}(i)u_{\la_2}(i)$ into what we shall define as $\tilde{I_1}$ 
and using Eq.~(\ref{eq:Gaussian_r}), we get,
\be
r=sr_0 + 12 v_0 s^\beta \tilde{I}_1 \;\;,
\ee
with the fixed point equation,
\be r =  { 12 v_o s^\beta \tilde{I}_1\over 1-s}\;\;. \label{eq:rRG1stO}\ee

The uncontracted term in $\langle V \rangle $ with all the fields in the 
lower-$\la$ range (Fig.~\ref{fig:feynmangraphs}b) gives, again by the same 
kind of power counting as above, a renormalized coupling constant
\be
v= v_0 s^{2\beta} \tilde{\Phi} \;\;,\label{eq:relevant}
\ee
 where possible corrections to scaling are included in $\tilde{\Phi}$.
  This equation has zero as 
fixed point, $v^\ast=0$, leading to $r^\ast=0$ via Eq.~(\ref{eq:rRG1stO}).  Thus, to 
first order in $v_0$, the Gaussian fixed point is 
stable, analogously to the Euclidean case.

We can now make a generalization, using our definition $ D-1 = -\beta $. 
Consider a Feynman diagram with $n$ legs. This will mean $n$ integrals 
over $\la$, in the absence of ``momentum conservation," and therefore a 
factor of $s^{-nD}$.  Each leg carries a factor of $\hpsi^<$, and 
therefore 
we get a factor of $z^n=s^{(D+1)n/2}$, resulting in a re-scaling factor of 
$s^{(1-D)n/2}=s^{n\beta/2}$.  But $\beta>0$ and the power of $s$ is now 
always positive. We see that there is no analogue of the upper critical 
dimensionality, above which the $\psi^4$ interaction terms become 
irrelevant in the ordinary $\psi^4$ Euclidean theory for scalar fields. 
(One recovers the Ginzburg-Landau-Wilson result for $n=4$ by substituting 
$d$ for $D$, taking only $n-1$ factors of $s^{-D}$ and  $z=s^{(d+2)/2}$ due 
to the $q^2$ diffusive term, rather than a linear term in $\la$.)

All the contributions to the renormalized Hamiltonian, to all orders in 
perturbation theory, will grow indefinitely under successive renormalizations, 
i.e., are ``relevant" in the RG terminology. To first order, we are only saved by 
the fact that there exists a unique fixed point at $v^\ast=0$. We seem to be faced 
with an uncontrollable phenomenon of ``proliferation" !

\subsection{Second order in perturbation theory}

Now we would like to see explicitly what happens when we take into account terms that are higher order in $v_0$. 

The bubble diagram 
(Fig.~\ref{fig:feynmangraphs}c) is already familiar from the Euclidean case, 
and contributes to the renormalized 4-vertex. Using the results of  Appendix B,  we have

\be
 v\,= \,v_0 s^{2\beta}  \left[ \tilde{\Phi} - 36 v_0  (I_1)^2\right]\;\;.\label{eq:v_renorm}
\ee

There are further terms which contribute to the quadratic coupling, which would not survive in the ordinary case because of the ``momentum conservation" rule ( two ``lower"  momenta cannot be summed with two ``upper" ones to give zero).  These calculations are given in Appendix B, and here we will only  report the results.

Performing the rescalings over the second order contributions and putting together all the results, we have,
\be
r\,= \,sr_0 +12 v_0 s^\beta I_1- v_0^2\left[ (72+48) (I_1)^3 \right] s^\beta\;\;.
\ee

The solution for the fixed point equations, besides the trivial one where $r^\ast=0,\; v^\ast=0$,  are
\be
v^\ast = { \tilde{\Phi} -s^{-2\beta} \over  (I_1)^2}\;\;, \label{eq:fp1}\ee
and
\be r^\ast = s^\beta v^\ast I_1{12- v^\ast \left[ 120\, (I_1)^3\right] \over 1-s}\;\;. \label{eq:fp2}\ee

Taking $I_1$ to lowest order in $r_0$ gives,
\be
I_1 \sim \Lambda^{-\beta} {(s^\beta -1) \over \beta}
\ee
which leads to 
\be v^\ast \propto \frac{\beta^2 (1-s^{-2\beta})\Lambda^{2\beta}}{(s^\beta-1)^2}\;\;.\label{eq:fp3}
\ee
We see that, up to the approximations we have made, the $s$ dependence does not drop out 
of the Eqs.~(\ref{eq:fp1},\ref{eq:fp2}). 
Apart from the fixed point at $v^\ast=0$, there is no finite fixed point for $v$. 

There is one other connected graph to second order in $v_0$, which is obtained by
taking one field from each subgraph and contracting them (see 
Fig.~\ref{fig:feynmangraphs}f) obtaining a $\psi^6$ 
interaction. We have computed its scaling factor explicitly, and find, in 
accordance 
with the power counting scheme of the previous section, that it scales like $s^{3\beta}$, i.e., it is yet another relevant coupling. 

The Laplace spectrum for a scale free network, shown in Figs.~\ref{fig:BAddLS},\ref{fig:BALStail}, is not a homogeneous function over the entirety of its range. We could think of taking  the re-scaling parameter to be very close to unity, $s=1+\delta$ with $\delta\ll 1$ but
this does not help us find a non-trivial fixed point, and taking 
$\delta\to 0$ leads to a blow-up of Eq.~(\ref{eq:fp3}), just as taking $\Lambda\to 
\infty$ does. Substituting (\ref{eq:fp3}) into (\ref{eq:fp2}) leads to the same problems.
We must conclude that the only physical fixed point in this perturbative treatment is at $r^\ast=0,\; v^\ast=0$.

\section{Stochastic and deterministic complex networks}
\label{sec:stochastic}

\subsection{Replica approach to a quenched average over the spectral density}
\label{ssec:average}

In the previous subsections we have pushed forward the computations with the naive assertion that the spectral density scales simply as $\rho(s\la)\sim s^{-\beta} \rho(\la)$. However, it is clear from the Figs. \ref{fig:BAddLS} and \ref{fig:BALStail} that the truth is somewhat more complicated than that. In particular, we proposed that  the average over 100 realizations of the network can be represented by the Weierstrass-Mandelbrot function (\ref{eq:WMfunction}). Therefore, in order to really speak about a non-stochastic spectral density, we should take an average over different realizations.  In the present case, this has to be a quenched average taken over the free energy of the system, since by assumption the network is fixed and does not fluctuate within the relaxation times of the fields (spins) living on the network. 

It is standard to use the ``replica method"~\cite{Kac,Edwards,EA} to be 
able to take the average over the logarithm of the partition function. The 
trick is to introduce $n$ independent replicas of the system and to take 
the average over $ Z^n$, finally using the identity $\lim_{n\to 0} 
(Z^n-1)/n = \ln Z$.

To illustrate how the calculations have to be done let us take the non-interacting, Gaussian theory as a starting point. Then, 
\be
 Z^n = \int_{-\infty}^\infty \prod_{\alpha=1}^n \prod_z d\hpsi_{z}^{(\alpha)} e^{-\sum_\alpha H_0^\alpha}\;\;,
\ee
where $\alpha=1,\ldots n$ is the replica index,
\be
  \sum_{\alpha=1}^n H_0^\alpha = \frac{1}{2} \int_0^\Lambda dz {1 \over N}\sum_i^N\sum_\alpha^n 
\hpsi_z^{(\alpha)}(r_0+z)\hpsi_z^{(\alpha)}\delta(z-\lambda_i)\;\;.  \label{eq:replicas}
\ee
and we have used Eq. (\ref{eq:rho}).  

If the distribution over different realizations of the spectrum $\la_0,\ldots,\la_{N-1}$ is denoted by 
${\cal P} (\{\la_i\}) $, we have 
\be
\langle Z^n\rangle = \int_{-\infty}^\infty \prod_{\alpha=1}^n 
\prod_z d\hpsi_z^{(\alpha)}  
\int d\la_0\ldots d\la_{N-1} {\cal P} (\{\la_i\}) e^{-\sum_\alpha H_0^\alpha}\;\;.
\ee
This expectation value now looks like a path integral over individual paths $\{\la_i\}$. Let us define
\be
\langle {1\over N} \sum_{i,\alpha} (r_0+z) \delta(z-\la_i)\rangle \equiv \rho(z)\;\;.
\ee
A cumulant 
expansion up to second order gives,  
\be  \langle e^{-\sum_\alpha H^\alpha_0} \rangle \simeq e^{-C_1 + {1\over 2} C_2}\;\;\; \ee
where 
\bea
\lefteqn{C_1 = {1\over 2} \langle \int_0^\Lambda  dz {1\over N} \sum_{i,\alpha} (r_0+z) 
\delta(z-\la_i) (\hpsi_z^\alpha)^2 \rangle =} \nonumber \\
& & \mbox{} \sum_{\alpha} \int dz (r_0 +z) \rho (z) (\hpsi_z^\alpha)^2 
\;\;.
\eea
 The second cumulant gives rise to 
bi-quadratic terms which couple different replicas,
\bea
\lefteqn{C_2 = {1\over 4} \int_0^\infty \prod_i d\lambda_i {\cal P} (\{\la_i\})\int dx dy {1 \over N^2}} \nonumber \\ 
& &\times \sum_{i,j}\sum_{\alpha,\beta} \left[(r_0+x) (r_0+y)\; \delta (x-\la_i) 
\; \delta (y-\la_j)\right] (\hpsi_x^\alpha)^2 
(\hpsi_y^\beta)^2  - C_1^2 \nonumber\\
& & \mbox{} = {1\over 4}\int dz (r_0+z)^2 \,\rho(z) \, \sum_{\alpha,\beta} (\hpsi_z^\alpha)^2 (\hpsi_z^\beta)^2\} \;\;,
\eea
where the off-diagonal terms in $i,j$ have been cancelled by $C_1^2$.
Thus it turns out that averaging the spectral density over different realizations, one ends up with a bi-quadratic term in the Hamiltonian, which could have been included  from the start. 
We get, for the effective Hamiltonian, 
\bea 
\lefteqn{H^{\rm eff}=
{1\over 2} \int_0^{\Lambda} d\la \rho(\la)\; (r_0+ \la)\; \sum_\alpha 
[\hpsi^{(\alpha)}_\la]^2 } \nonumber \\ 
& &  \mbox{} - {1\over 4} \int_0^{\Lambda} d\la \rho(\la )(r_0+\la)^2 
\sum_{\alpha,\beta}[\hpsi^{(\alpha)}_\la]^2[\hpsi^{(\beta)}_\la]^2 \;\;.
\eea
Note, however that there is no small parameter which allows us to terminate the cumulant expansion or in which to expand the biquadratic
term. (For treatment of this problem via functional RG methods see~\cite{Zinn-Justin}.

The first term in the Hamiltonian, up to the summation over the replica indices, is essentially what we started off with in 
Section~\ref{ssec:naivegaussian} and in the $n\to 0$ limit will give the same free energy.  The renormalization of the quadratic term will therefore be the same, with $r=sr_0$.  The bi-quadratic terms with the new interaction  
$U_4\equiv r_0^2+2 r_0 \la +\la^2$  now have to be treated in a perturbative fashion relative to the Gaussian term, using another cumulant expansion.
The different terms in the uncontracted vertex acquire the re-scaling factors $s^{3-\beta}$, $s^{2-\beta}$ and $s^{1-\beta}$,  in this order.  
Although this has to be checked in detail, we conjecture that they are thus irrelevant for $\beta> 3$, and therefore to first order the theory remains Gaussian. The new 
quadratic couplings coming from the once-contracted graphs, however, depend on $\lambda$ and are not trivial. 

 Inclusion of a $v_0\psi^4$ interaction, under the quenched averaging, 
gives rise to a bi-quadratic and a quartic term where the number of $\la$ 
integrals are reduced to 2 and 1 respectively, with further 
simplifications now due to Gaussian contractions in the cumulant expansion 
in $v_0$.  Power counting suggests that these terms are irrelevant for 
$\beta> 3/2$ and $\beta > 5/3$ to first order in $v_0$.

\subsection{Deterministic hierarchical lattices and the matrix extension method}

For some hierarchical networks obtained by successive decorations of a seed graph, 
the spectrum of the normalized Laplacian can be computed iteratively.\cite{Molazemov}
Under graph decoration, the``matrix extension transformation" yields all the 
new eigenvalues, in terms of the existing ones~\cite{Teplyaev,Zhang}.
It has been shown that most of the spectrum is given by the 
pre-images of the so called ``spectral decimation" transformation $ \lambda = 
R(\lambda^\prime )$, and converges to the Julia set of $R^{-1}$ as $N\to 
\infty$~\cite{Molazemov}. Moreover,  one can easily check  that the attractor (the Julia set) is chaotic, with the preimages of many eigenvalues jumping back and forth between different intervals  $0< \lambda <1$ and $\lambda >1$.  We conjecture that similar properties may also hold for the un-normalized Laplace spectrum which we have been treating in this paper.  Thus there is no smooth way in which to rescale existing eigenvalues to restore the integrated-out ones; although the network is deterministic, the rescaling transformation on the spectrum itself could just as well be stochastic.

It is interesting to recall a chaotic ``real space" renormalization group 
transformation~\cite{Derrida_chaoticRG} encountered in a frustrated Potts 
model on a hierarchical lattice.  This example shows that a finite 
interval below the critical temperature may become densely populated with 
critical points (singularities of the free energy) leading to BKT like 
behavior, similar to that found by Andrade and 
Herrmann~\cite{Andrade_RSRG}.

\section{Conclusions}
\label{sec:conclusions}

In this paper we have tried to explore the possibilities offered by the eigenvectors and eigenvalues of the graph Laplacian to develop a field theoretic renormalization 
group (FTRG) approach to order-disorder phenomena on complex networks. We have taken a tutorial approach which intends to acquaint the reader with 
the ideas of FTRG, and then to build upon these ideas in order to develop the analogous machinery on a complex network. We have been able to carry over most of the basic 
concepts and to implement many of the usual procedures.

The proliferation of higher order terms in the renormalized Hamiltonian was brought under control by going over to a quenched average over different realizations of the stochastic 
network using a replica approach. We found that for $\beta>3$, the first order perturbation expansion in the bi-quadratic couplings introduced by the quenched average were 
irrelevant, and that the theory remained Gaussian. Inclusion of a quartic interaction can also be shown to be irrelevant for $\beta>3/2$ up to first order in $v_0$.  Further work is in progress.

The $\la$-dependence acquired by the effective renormalized couplings persist, and in principle would call for 
a nonlinear transformation $\lambda^\prime = T_s(\lambda)$ 
parameterized by the scale factor $s$.  This, in turn,  calls for an initial 
$\lambda$-dependent temperature like coupling constant, $r(\lambda)$. 
The $\lambda$-dependent transition temperature, also indicated by the calculation in section~\ref{ssec:average}, may be a way to understand the Berezinskii-Kosterlitz-Thouless (BKT)  power law behavior of correlations obtained over a range of temperatures for $T<T_c$ found in Real 
Space RG calculations on scale free networks.~\cite{Andrade_RSRG}

\ack

AE would like to thank Zehra \c Cataltepe for having introduced her to the graph Laplacian, Lucilla Arcangelis and Anita Mehta for a careful reading of the manuscript. She also acknowledges partial support by the Turkish Academy of Sciences.

\appendix
\section{Normalized Laplacian} \label{sec:AppA}

It is possible to define the  ``normalized" graph Laplacian via
\be
\tilde{\bf L}\,\equiv {\bf D}^{-1}{\bf L} = \, {\bf I} - {\bf D}^{-1}{\bf A}\;\;,
\ee 
where ${\bf I}$ is the identity matrix.  

Eigenvectors of the normalized Laplacian are no longer orthogonal to each 
other, but one has,
\be {\bf v}_\mu^\dagger {\bf D} {\bf v}_{\mu^\prime}=\delta_{\mu, 
\mu^\prime}\;\;.
\ee
An advantage to using the normalized Laplacian in this context is that the 
eigenvalues are confined to a fixed interval $[0, 2]$ such that
$\mu_0=0$, $0 <\mu_1 \le \mu_2 \ldots \mu_{N-1} \le 2$.  In this way 
the eigenvalue spectrum gets denser and smoother as the graph becomes 
larger.

For many properties of normalized and unnormalized Laplacian spectra, see 
Ph. D. thesis of A. Banerjee.~\cite{Banerjee}
 
\section{Second order contributions to the renormalized Hamiltonian}\label{sec:AppB}

For compactness of notation  we have defined $d_\mu \equiv d \la_\mu \rho(\la_\mu)$, where $\la_\mu$ appears in a subscript it will be just represented by $ \mu$, and 
$\int_{\Lambda/s}^\Lambda \equiv \int^>$.

\begin{figure}[ht]
\begin{center}
\includegraphics[width=6cm]{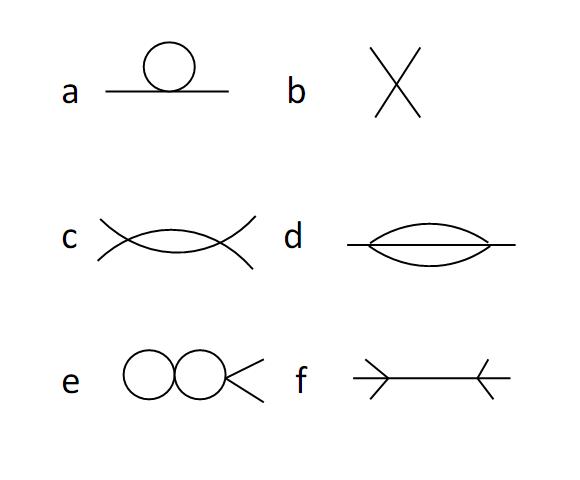}
\end{center}
\caption{\footnotesize The Feynman diagrams to which reference is made in the text. In the context of the $\psi^4$ theory, the diagrams a), e) and d) contribute to the 
renormalized quadratic couplings; b) and c) to the quartic coupling, and f) to a sixth order coupling. As usual, a vertex (where four lines meet) is proportional to the 
coupling constant $v_0$, the 
``internal" lines (which are connected to a vertex at both ends) denote the Green's function $G(r_0,\la)$, Eq. (\ref{eq:green}), and each ``external" leg carries a factor 
$\hpsi_\la^{\rm lower}$.
}
\label{fig:feynmangraphs}
\end{figure}

The second order contribution from the perturbation expansion (Fig.~\ref{fig:feynmangraphs}c)  to the $\psi^4$ interaction in the renormalized Hamiltonian is given by  the doubly contracted term,
\be 
36 v_0^2\int^<\ldots \int^< \prod_{\mu=1,\ldots,4} \left[\, d_\mu \hpsi^<_\mu\,\right] Q_4^{(2)} 
\ee
and  
\be
Q_4^{(2)} =  \sum_{i,j}  \{u_1(i) u_2(i) u_3(j) u_4(j)  \int^> \int^> \prod_{\nu=a,b} \left[ d_\nu G(\la_\nu)\right] u_a(i) u_a(j) u_b(i) u_b(j) \} 
\ee
where we have re-numbered  lines so that the indices $\mu=1,2,3,4$ belong to the external legs and  $\nu=a,b$ belong to the internal lines; $i,j$ index the elements of the eigenvectors. Defining, after Eq. (\ref{eq:I_1}),
\be
I_2(i,j) \equiv \int^> d\la \rho(\la)\, G(\la)\, u_\la(i) u_\la(j)\;\;,
\ee 
we may re-write,
\be
Q_4^{(2)} =  \sum_{i,j}  \{ u_1(i) u_2(i) u_3(j) u_4(j)\left[ I_2(i,j) \right]^2 \} \;\;.
\ee
The effective coupling  $36v_0^2 Q_4^{(2)}$ depends on the eigenvectors $\lambda_1, \ldots \la_4$. 
 The re-scaling factor for this 4-vertex can be obtained by power counting to be $s^{2\beta}$, up to the corrections which might 
come from this dependence.  In the best RG tradition~\cite{Goldenfeld} we will neglect the $\la$ dependence of the 4-vertex in what follows. If we also neglect the $(i,j)$  dependence of $I_2$ and $I_1$ (which then become identical) we simply have
\be Q_4^{(2)} \propto  (I_1)^2\;\;.
\ee
Expanding $I_1$ in $r_0$ gives, from Eq. (\ref{eq:I_1}),
\be
I_1 =\simeq \Lambda^{-\beta} \left[ {1\over \beta} (s^\beta -1) -r_0 {\Lambda^{-1}\over \beta+1} (s^{\beta+1}-1)\right]\;\;.
\ee

One contribution (Fig.~\ref{fig:feynmangraphs}d) the quadratic coupling to the second order is, with 1,2 denoting the external legs,  
\begin{eqnarray}
\lefteqn{ Q_{2,1}^{(2)}=72  v_0^2   \sum_{i,j}  u_1(i) u_2(i)  
 \int^> \int^>\int^> \prod_{\mu=3,4,5} d_\mu G(\la_mu)  u_3(i) u_3(j)u_4(i) u_4(j) u_5(j)^2 } 
\nonumber \\
& & \mbox{} = 72 v_0^2 \sum_{i,j} u_1(i) u_2(i)\left[ I_2(i,j)\right]^2 I_1(i) \;\;,\label{eq:Q21^2}
\end{eqnarray}
Note that, to the same approximation as above, we have 
\be Q_{2,1}^{(2)} \propto (I_1)^3\;\;.\label{eq:Q21app}\ee 
The other contribution is given by  is (Fig.~\ref{fig:feynmangraphs}e)
\be
Q_{2,2}^{(2)} = 48 v_0^2 \sum_{i,j} u_1(i) u_2(j)\left[ I_2(i,j)\right]^3 
 \propto (I_1)^3\;\;.\label{eq:Q22app}\ee


\end{document}